\documentclass[aps,prl,floatfix,twocolumn]{revtex4}

\usepackage{graphicx}
\usepackage{epsfig}

\usepackage{url}
\usepackage{paralist}
\usepackage{textcomp}

\begin{document}

\begin{widetext}
\noindent\textbf{Preprint of:}\\
Simon J. Parkin, Gregor Kn\"{o}ner, Timo A. Nieminen,
Norman R. Heckenberg and Halina Rubinsztein-Dunlop\\
``Picolitre viscometry using optically rotated particles''\\
\textit{Physical Review E} \textbf{76}(4), 041507 (2007)
\end{widetext}

\title{Picolitre viscometry using optically rotated particles}

\author{Simon J. Parkin}
\author{Gregor Kn\"{o}ner}
\author{Timo A. Nieminen}
\author{Norman R. Heckenberg}
\author{Halina Rubinsztein-Dunlop}

\affiliation{Centre for Biophotonics and Laser Science, School of
Physical Sciences, The University of Queensland, QLD 4072,
Australia}


\begin{abstract}
Important aspects in the field of microrheology are
the studies of the viscosity of fluids within structures with
micron dimensions and fluid samples where only microlitre
volumes are available. We have quantitatively investigated the
performance and accuracy of a microviscometer based on rotating
optical tweezers, that requires as little as one microlitre of
sample. We have characterised our microviscometer, including
effects due to heating, and demonstrated its ability to perform
measurements over a large dynamic range of viscosities (at
least two orders of magnitude). We have also inserted a probe
particle through the membrane of a cell and measured the
viscosity of the intramembranous contents. Viscosity
measurements of tears have also been made with our
microviscometer, which demonstrate its potential use to study
un-stimulated eye fluid.
\end{abstract}

\maketitle

Recently, there has been increasing interest in microrheology,
the study of flows and deformations of a material or medium
using probes of microscopic size. In this paper we will
concentrate on microrheological methods that probe viscosity on
micrometre length scales. Suitable existing techniques are
magnetic tweezers \citep{crick1950,bausch1999}, particle tracking
\citep{mason1997} and optical tweezers based techniques
\citep{starrs2002,bishop04}. Magnetic tweezers allow
comparatively large forces to be applied to probe particles and
thus the effects of high rates of shear to be studied. Particle
tracking elegantly extracts the viscoelasticity of a medium
over a large frequency range and allows fluid-probe coupling
effects to be removed \citep{crocker2000}. Optical tweezers
allow the viscoelasticity of very localised regions to be
probed, which enables the investigation of femtolitre volumes
and micrometer structures, such as the interior of cells. The
region can be further localised by studying rotational motion
of the trapped particle \citep{bishop04}, which is also true for passive techniques \citep{cheng2003,andabloreyes2005prl,schmiedeberg2005el}. These techniques have
been used to study the viscoelasicity of cells
\citep{valberg1985,daniels2006} and also polymer solutions,
where small volumes and high throughput are advantageous
\citep{breedveld2003}. Another potential application is small
volume medical samples, such as eye fluid \citep{tiffany1991}.

Rotating optical tweezers have been discussed in detail by
Parkin \textit{et al.} \citep{parkin2007mcb}. A spherical
birefringent microparticle, combined with an optical
measurement of the torque applied to it, can be used to probe
fluid properties \citep{nieminen2001,bishop04}. Using the light
transmitted through the probe particle trapped in optical
tweezers, the rotation rate of the probe particle and the
change in the polarisation of the light are measured. We use
vaterite, which is a calcium carbonate crystal that forms
spherical structures under certain growth conditions
\citep{bishop04}, as our probe particle. This particle has also
been used to create and study microfluidic flows
\citep{knoener2005,leach2006,dileonardo2006}. We present the
characterisation of our microviscometer, based on this rotating
sphere, and the application of this device to measure
intramembranous liquid and tear fluid. This technique allows a flow to be generated in a very localized region, of picolitre volume, and the viscosity of the fluid in this region to be measured. The use of rotational motion means that a smaller volume is probed compared to methods based on translational motion, due to the tighter confinement of the flow. This demonstrates the potential of this method as a high-resolution active-probe method for microviscometry.

The optical torque applied to a birefringent sphere by the
trapping laser is \citep{bishop04}:
\begin{equation}
\tau_{\textrm{optical}} = \frac{\Delta\sigma P}{\omega}
\label{eq:opticaltorque}
\end{equation}
where $\Delta \sigma$ is the change in the degree of circular
polarisation as the beam passes through the particle, $P$ is the
laser power and $\omega$ is the optical angular frequency.
Viscosity is found by equating the applied torque and the viscous
drag torque on a rotating sphere. The drag torque is complicated
by the fact that experimentally we show that the viscosity varies
with the trapping laser power, which is explained by heating of
the fluid due to slight absorption of the trapping laser by the
probe sphere. This leads to a non-uniform temperature distribution
within the liquid. The steady state temperature of the fluid
around a sphere, that has a fixed uniform surface temperature, as
a function of the distance from the centre of the sphere, $r$, is
given by:
\begin{equation}
T(r) = \frac{\gamma}{r} + T_0 \label{eq:temp_r}
\end{equation}
where $\gamma$ is a constant and $T_0$ is room temperature. The
viscosity of a fluid varies with temperature which means there
will be a non-uniform distribution of fluid viscosity around
the sphere. For certain liquids, experimental data exists which
can be used to determine viscosity from temperature
\cite{barnes1989}. Theoretical models for viscosity as a
function of temperature tend to be inaccurate over significant
temperature ranges, so we use interpolated data to determine
viscosity as a function of temperature.

\begin{figure}
\begin{center}
\includegraphics{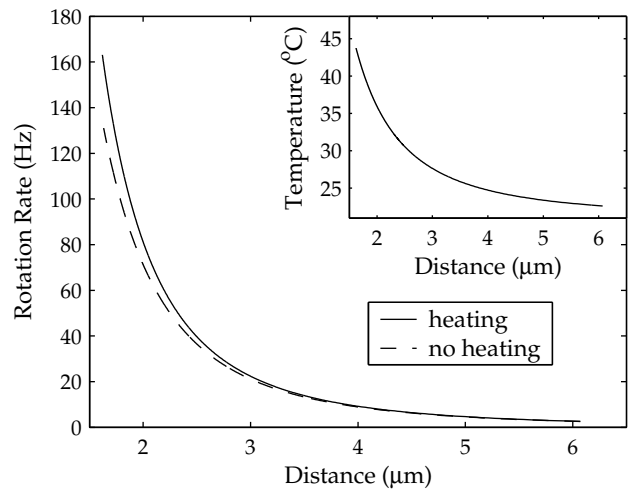}
\end{center}
\caption{Rotation rate of the fluid in the equatorial plane as a
function of the distance from the centre of the particle. The
rotation rates correspond to a particular optical torque from
an experiment with a vaterite particle, $3.2$\,\textmu m in
diameter. The solid line represents the case where the fluid
has a temperature variation as shown in the inset. The
temperature is constant throughout the fluid for the dashed
line.}
\label{fig:shellrotation}
\end{figure}

For steady state creeping flow, in an infinite viscous medium,
driven by a rotating sphere, the fluid flow at any radius can
be characterised by an angular velocity. As a torque must be
applied to the rotating sphere to maintain the flow against
viscous drag, there is a uniform outward flux of angular
momentum equal to \citep{landau1987}:
\begin{equation}
\tau = 8 \pi \eta(r) r^6 \frac{d\Omega}{d(r^3)}
\label{eq:dragtorque}
\end{equation}
where $\eta$ is the viscosity of the surrounding liquid and
$\Omega$ is the angular frequency of rotation of the sphere. This
is the case even when the viscosity is non-uniform as long as its
distribution within the fluid is spherically symmetric. The
rotation rate of the fluid is equal to the rotation rate of the
particle, at the particle's surface, which we experimentally
measure and is given by:
\begin{equation}
\Omega = \frac{\tau}{8 \pi} \int_{r=\infty}^{r=a}\frac{1}{\eta(r)}
d(1/r^3) \label{eq:rotationrate}
\end{equation}
We do not have an analytical expression for $\eta(r)$, however
in this form the integral can be easily calculated numerically.
The torque, $\tau$, in this equation is equal to the optically
applied torque, given by equation \ref{eq:opticaltorque} and is
measured experimentally. The surface temperature of the
particle is unknown and is required to determine $\eta(r)$. The
surface temperature, according to equation~\ref{eq:temp_r},
depends on the parameter $\gamma$. However, from the empirical
relation of viscosity as a function of temperature, which is
derived from tabulated experimental data, and the relationship
between temperature and distance, r, from
equation~\ref{eq:temp_r}, it is possible to rewrite
equation~\ref{eq:rotationrate} as:
\begin{equation}
F(\gamma) = 0 \label{eq:gammafunction}
\end{equation}
which can be numerically solved using the Newton--Raphson
method to find $\gamma$. The rotation rate of the fluid shells
in this model, as a function of distance from the centre of the
particle, is shown in figure \ref{fig:shellrotation}. It can be
seen that although the heating effect is quite localised there
can be a significant effect on the rotation rate and hence, the
estimated viscosity.

\begin{figure}
\begin{center}
\includegraphics{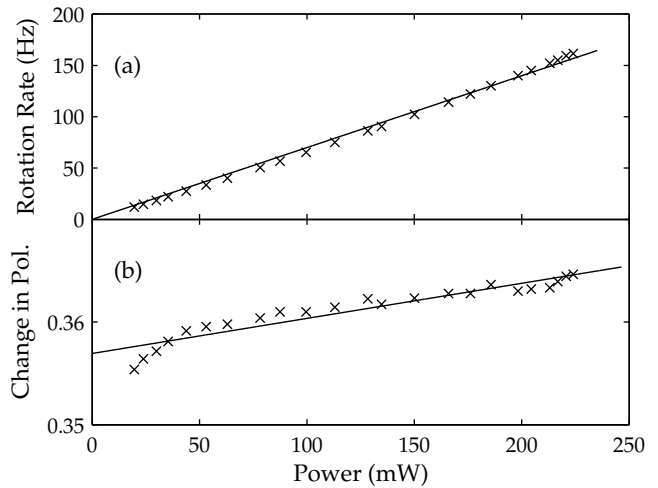}
\end{center}
\caption{The variation of rotation rate (a) and change in polarisation
(b) with trapping laser power. The data is for a vaterite
particle with a diameter 3.2\,\textmu m in methanol. The errors
in these plots are small enough that the trends are significant.
The absolute value of the power is not known precisely but the
relative values of the powers in these plots are, and are
limited only by the detector's precision. The rotation rate is
precise, provided enough revolutions of the particle are
recorded, which means the deviation from linearity is
significant. The error in the polarisation measurement is
$0.005$, yet the scatter is less than this which means the
linear decrease with power is a real effect.}
\label{fig:RotationPolarisation}
\end{figure}

The experimental setup used for this experiment is described in
\citep{bishop04}, and in more detail in \citep{parkin2005}. To
characterise the viscometer the power dependence of the
rotation rate and polarisation were determined (figure
\ref{fig:RotationPolarisation}). A linear fit, through the
origin, of the rotation rate as a function of power is shown
(figure \ref{fig:RotationPolarisation}(a)). Contrary to
expectation, the dependence is not perfectly linear which
suggests that another parameter varies with the laser power.
There is also unexpected behaviour of the change in
polarisation as a function of power (figure
\ref{fig:RotationPolarisation}(b)), which varies by several
percent over the range of powers measured. In this case the
dependence seems to be linear and a linear fit of the data is
shown. It is possible that this trend could be caused by
convection caused by heating of the fluid surrounding the
sphere, or could be due to an increase in trap strength due to
increasing the trapping laser power. However, the effect is
minimal and would be difficult to confirm experimentally.
Therefore we have not investigated the trend in this paper.


\begin{figure}
\begin{center}
\includegraphics{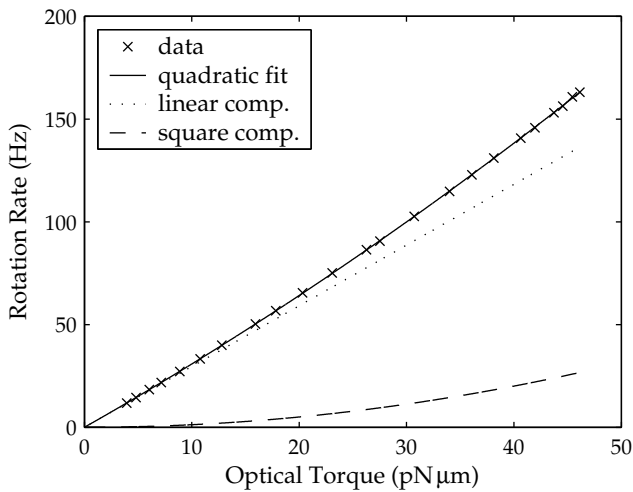}
\end{center}
\caption{The variation of rotation rate with the optically applied
torque. The components of the quadratic fit are shown. The
linear component is the expected rotation rate without heating
or non-Newtonian effects. The square component represents the
non-linear effects on viscosity, which in this case, are due to
heating that increases with laser power.}
\label{fig:opticaltorqueRR}
\end{figure}

\begin{figure}
\begin{center}
\includegraphics{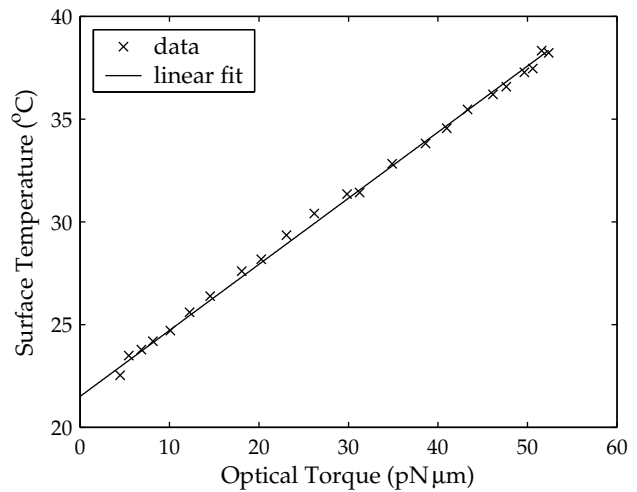}
\end{center}
\caption{Surface temperature of the vaterite particle in methanol due to heating from absorption of the trapping laser. This is the
maximum temperature in the surrounding liquid, the temperature
falls off as $r^{-1}$, which is depicted in
figure~\ref{fig:shellrotation}.}
\label{fig:surfacetemperature}
\end{figure}

Typically a viscometer measures viscosity by applying a known
or controlled stress to the medium of interest, and then
measures the resulting strain, which manifests as a shear rate
in a liquid. In our microviscometer the applied stress is
represented by the optically applied torque to the particle and
the shear rate is related to the rotation rate. Therefore the
rotation rate as a function of optical torque is the
relationship of interest, and is depicted in figure
\ref{fig:opticaltorqueRR}. The form of the fit is:
\begin{equation}
\Omega(\tau) = \alpha \tau + \beta \tau^2
\end{equation}
where $\alpha$ and $\beta$ are constants and $\tau$ is the
optically applied torque. The nonlinear response could either
be due to non-Newtonian behaviour of the fluid or a temperature
effect. In this case, as the fluid (methanol) is Newtonian,
there must be some heating occurring as the trapping laser
power is increased. Absorption of the laser light by the liquid
itself turns out to be of insufficient magnitude to explain the
decrease in viscosity \citep{peterman2003}. However, if the
particle is itself slightly absorbing, then the heating could
be of sufficient magnitude to explain the observed behaviour.
If this is the case, then the temperature of the fluid
surrounding the particle is described by equation
\ref{eq:temp_r}. Solving equation~\ref{eq:gammafunction} for
$\gamma$, gave a value for the rotation rate that matched the
experimentally observed rotation rate of the particle. The
surface temperature of the vaterite particle as a function of
optical torque was found and is plotted in figure
\ref{fig:surfacetemperature}. A linear fit of the surface
temperature data yields a temperature increase of
$66$\,$^\circ$C$/$W of laser power. Commonly used laser powers,
in our experiments and other optical tweezers experiments, are
of the order of 100\,mW, which corresponds to only a
$7$\,$^\circ$C temperature increase. The observed temperature
increase corresponds to 0.08\% of the laser power being
absorbed. Now that the power dependence of the measurements is
characterised and understood, accurate measurement of viscosity
of viscosity at room temperature is easily derived by making
measurements at several power levels, which is easily done, and
extrapolating to zero power. All the measurements discussed in
this paper were made in this way.

\begin{figure}
\begin{center}
\includegraphics{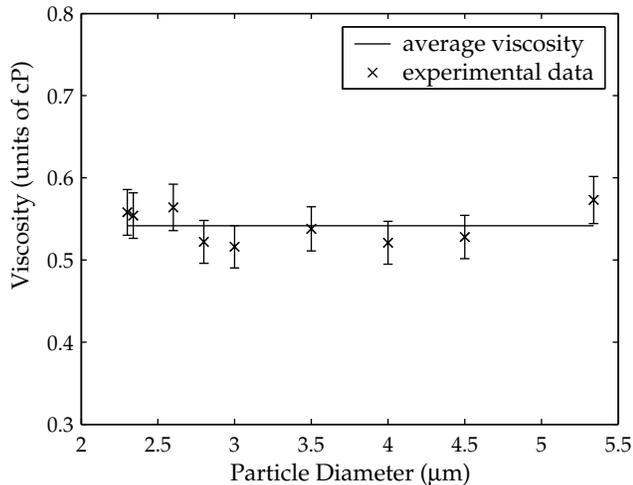}[th]
\end{center}
\caption{The viscosity of methanol measured using vaterite with
different diameters. This does not represent the the limit of
the size range of vaterite particles, as particles as small as
$1$\,$\mu$m and as large as $10$\,$\mu$m have been observed.
However, viscosity measurements were not carried out with these
spheres as the measurement of diameter become less accurate for
small spheres and the larger particles tend to be less
spherical.} \label{fig:particlesize}
\end{figure}

\begin{figure}[!ht]
\begin{center}
\includegraphics{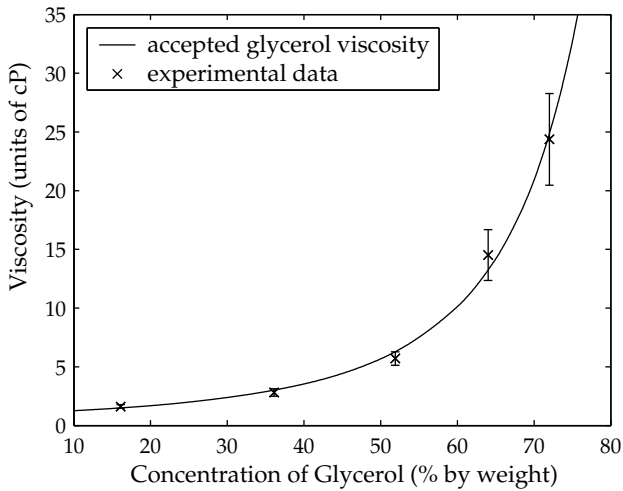}
\end{center}
\caption{Viscosity measurements for different concentrations of
glycerol. Each data point is the average of five measurements
using five different particles. The error bars represent the
standard deviation and the error in measurement of particle
diameter is primarily responsible for the data spread. The
relative error for each data point is about 10--15\%.}
\label{fig:glycerolviscosity}
\end{figure}

A good test of reproducibility is the microviscometer's
performance over a range of particle sizes. The result of this
test is plotted in figure \ref{fig:particlesize}. For probe
particles 2--5\, \textmu m in diameter, the viscosity
measurements are consistent with the expected independence of
viscosity on particle size.

We also investigated the performance of our microviscometer
over a range of viscosities. A series of solutions with
different concentrations of glycerol were chosen because the
relationship between viscosity and glycerol concentration has
been well characterised \citep{segur1951}. In addition the
glycerol polymers are short, which means the fluid exhibits
Newtonian behaviour. The viscosities of different
concentrations of glycerol, as measured by our microviscometer,
are shown in figure \ref{fig:glycerolviscosity} compared to
accepted values and show good agreement. The dynamic range
tested here was two orders of magnitude, however that does not
represent the limits of our technique. Measurements of both
lower and higher viscosities are possible, with the upper limit
being determined by any user imposed time restrictions on
acquiring an accurate rotation signal. It is important to note
that the volume of the sample used for these measurements was
10--15\,\textmu L, approximately one drop of fluid. Dried
vaterite particles were added to the sample using the tip of a
brass wire while the sample was on the microscope slide. This
`in situ' addition of vaterite is advantageous as it allows
volumes as small as 1\,\textmu L to be handled.


Now that the microviscometer is fully characterised, it can be
used for practical applications. An example of a medical
sample, where only microlitre volumes are available is eye
fluid. The viscosity of eye fluid has previously been measured
to be 3--4\,cP \citep{tiffany1991}. Without stimulating a tear
response, only about 1--5\,\textmu L of eye fluid can be
collected \citep{tiffany1991}. In a proof of principle
experiment, we measured the viscosity of a similar volume of
stimulated tears using our microviscometer. We found the
viscosity to be $1.1\pm0.1$cP. In our experiment the collection
procedure was safe but crude, so we expected the measured tear
fluid viscosity to be close to water (0.97\,cP at
$21.5^{\circ}$C). In the future, more quantitative studies
could be carried out by employing controlled and reproducible
eye fluid extraction procedures developed by ophthalmologists
\cite{pandit1999eer}.

\begin{figure}[th]
\begin{center}
\includegraphics[width=8.5cm]{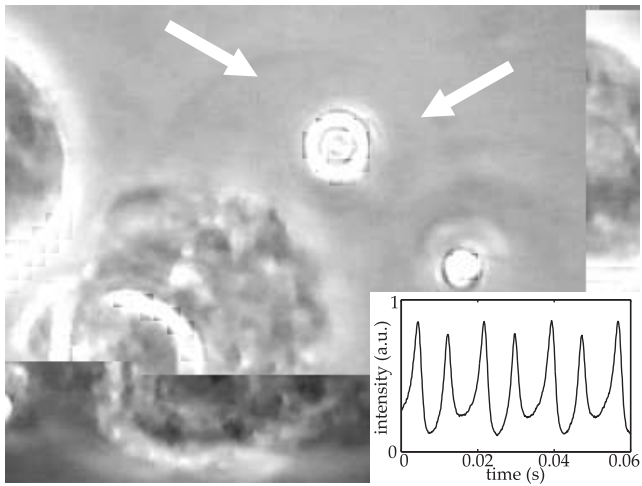}
\end{center}
\caption{A phase contrast image of a vaterite inside a cell. The left
arrow points to the cell membrane, or the `bleb'. The right
arrow points to the vaterite within the cell membrane. The
inset shows the normalised intensity, in arbitrary units, measured by the photo-detector used to determine the rotation rate of the vaterite particle.}
\label{fig:vateriteincell}
\end{figure}

Probing very localised regions of fluid is an application that
we have previously demonstrated by measuring the viscosity of a
fluid inside a micelle \citep{bishop04}. In that experiment the
vaterite particles were added during the formation of micelles
so that, on occasion, a vaterite particle was engulfed by a
micelle. A more interesting case is the viscosity inside a
cell. In a proof of principle experiment, we trapped and
rotated a vaterite particle within a `bleb' on a macrophage
cell (figure \ref{fig:vateriteincell}). The bleb, an extended
region of the cell membrane, was formed by the cell in response
to exposure to a focussed femto-second laser. The vaterite was
then inserted into the cell by simultaneously cutting a hole in
the cell membrane with a femto-second laser and pushing the
vaterite through the hole in the membrane using an optical
trapping laser beam. The viscosity was measured to be
$3.3\pm0.5$\,cP which suggests that the fluid was drawn into
the cell through the membrane during bleb formation, as the
surrounding fluid has a viscosity close to water while
intercellular viscosity has been reported to be orders of
magnitude higher \citep{daniels2006,valberg1985}.

We have demonstrated a microviscometer that measures viscosity by
optically applying torque to a spherical probe particle and
optically measuring both the torque and the particle's rotation
rate. The viscosities of glycerol solutions varying by two orders
of magnitude have been measured which demonstrates the lower limit
of the dynamic range. Effects due to absorption of the trapping
laser, which causes local heating, have been accounted for and
quantified. Experiments within cells and with eye fluid have
demonstrated the practical applications of this technique.


\end{document}